\documentclass[aps,prl,preprint,tightenlines,superscriptaddress,showpacs,byrevte
x]{revtex4}

\usepackage{graphicx} 
\usepackage{dcolumn}  
\usepackage{afterpage} 

\graphicspath{{ps}}

\renewcommand{\bar}[1]{\overline{#1}{}}

\draft 

\begin{document}

\vspace*{-3\baselineskip}
\resizebox{!}{3cm}{\includegraphics{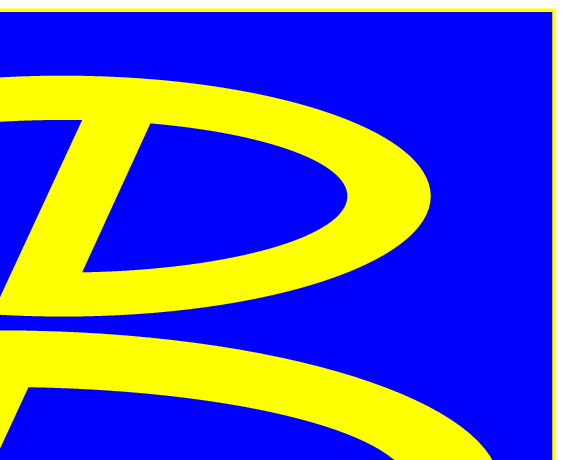}}

\preprint{\vbox{ \hbox{Belle Preprint 2006-12}}
\preprint{\vbox{ \hbox{KEK   Preprint 2006-6}}
}}

\title{ \quad\\[0.5cm] 
Measurement of $D^0\rightarrow\pi l\nu (K l\nu)$ Form
  Factors and Absolute Branching Fractions}

\tighten

\begin{abstract}
{Using a 282 fb$^{-1}$ data sample collected by the Belle experiment at
the KEKB $e^+ e^-$ collider, we study $D^0$ decays to $K^- \ell^+ \nu$ and $\pi^- \ell^+ \nu$ final states. The $D^0$ flavor and momentum are tagged through a full
reconstruction of the recoiling charm meson and additional mesons from fragmentation.
The reconstruction method provides very good resolution in neutrino momentum and in $q^2 = (p_\ell+p_\nu)^2$.
Normalizing to the total number of $D^0$ tags, we measure the absolute branching fractions to be
$\mathcal{B}(D^0 \rightarrow K \ell \nu)$ =($3.45\pm0.07_{\mbox{stat}}\pm0.20_{\mbox{syst}}$)\% and  
$\mathcal{B}(D^0 \rightarrow \pi \ell \nu)$ = ($0.255\pm0.019_{\mbox{stat}}\pm0.016_{\mbox{syst}}$)\%
and the semi-leptonic form factors (within the modified pole model) 
$f_+^K(0) = 0.695 \pm 0.007_{\mbox{stat}} \pm 0.022_{\mbox{syst}}$ and $f_+^\pi(0) = 0.624 \pm 0.020_{\mbox{stat}} \pm 0.030_{\mbox{syst}}$.}
\end{abstract}
 
\affiliation{Budker Institute of Nuclear Physics, Novosibirsk}
\affiliation{Chiba University, Chiba}
\affiliation{University of Cincinnati, Cincinnati, Ohio 45221}
\affiliation{University of Frankfurt, Frankfurt}
\affiliation{University of Hawaii, Honolulu, Hawaii 96822}
\affiliation{High Energy Accelerator Research Organization (KEK), Tsukuba}
\affiliation{Institute of High Energy Physics, Vienna}
\affiliation{Institute of High Energy Physics, Protvino}
\affiliation{Institute for Theoretical and Experimental Physics, Moscow}
\affiliation{J. Stefan Institute, Ljubljana}
\affiliation{Kanagawa University, Yokohama}
\affiliation{Korea University, Seoul}
\affiliation{Kyungpook National University, Taegu}
\affiliation{Swiss Federal Institute of Technology of Lausanne, EPFL, Lausanne}
\affiliation{University of Ljubljana, Ljubljana}
\affiliation{University of Maribor, Maribor}
\affiliation{University of Melbourne, Victoria}
\affiliation{Nagoya University, Nagoya}
\affiliation{Nara Women's University, Nara}
\affiliation{National Central University, Chung-li}
\affiliation{National United University, Miao Li}
\affiliation{Department of Physics, National Taiwan University, Taipei}
\affiliation{H. Niewodniczanski Institute of Nuclear Physics, Krakow}
\affiliation{Nippon Dental University, Niigata}
\affiliation{Niigata University, Niigata}
\affiliation{Nova Gorica Polytechnic, Nova Gorica}
\affiliation{Osaka City University, Osaka}
\affiliation{Osaka University, Osaka}
\affiliation{Panjab University, Chandigarh}
\affiliation{Peking University, Beijing}
\affiliation{Princeton University, Princeton, New Jersey 08544}
\affiliation{University of Science and Technology of China, Hefei}
\affiliation{Seoul National University, Seoul}
\affiliation{Shinshu University, Nagano}
\affiliation{Sungkyunkwan University, Suwon}
\affiliation{University of Sydney, Sydney NSW}
\affiliation{Tata Institute of Fundamental Research, Bombay}
\affiliation{Toho University, Funabashi}
\affiliation{Tohoku Gakuin University, Tagajo}
\affiliation{Tohoku University, Sendai}
\affiliation{Department of Physics, University of Tokyo, Tokyo}
\affiliation{Tokyo Institute of Technology, Tokyo}
\affiliation{Tokyo Metropolitan University, Tokyo}
\affiliation{Tokyo University of Agriculture and Technology, Tokyo}
\affiliation{University of Tsukuba, Tsukuba}
\affiliation{Virginia Polytechnic Institute and State University, Blacksburg, Virginia 24061}
\affiliation{Yonsei University, Seoul}
   \author{L.~Widhalm}\affiliation{Institute of High Energy Physics, Vienna} 
   \author{K.~Abe}\affiliation{High Energy Accelerator Research Organization (KEK), Tsukuba} 
   \author{K.~Abe}\affiliation{Tohoku Gakuin University, Tagajo} 
   \author{I.~Adachi}\affiliation{High Energy Accelerator Research Organization (KEK), Tsukuba} 
   \author{H.~Aihara}\affiliation{Department of Physics, University of Tokyo, Tokyo} 
   \author{K.~Arinstein}\affiliation{Budker Institute of Nuclear Physics, Novosibirsk} 
   \author{Y.~Asano}\affiliation{University of Tsukuba, Tsukuba} 
   \author{T.~Aushev}\affiliation{Institute for Theoretical and Experimental Physics, Moscow} 
   \author{A.~M.~Bakich}\affiliation{University of Sydney, Sydney NSW} 
   \author{V.~Balagura}\affiliation{Institute for Theoretical and Experimental Physics, Moscow} 
% ...
\collaboration{The Belle Collaboration}

\maketitle
\tighten
{\renewcommand{\thefootnote}{\fnsymbol{footnote}}}
\setcounter{footnote}{0}


Exclusive semileptonic decays of $B$ and $D$ mesons are a favored means of
determining the weak interaction couplings of quarks within the standard model
because of their relative abundance and simplified theoretical treatment.
The latter, given leptons are insensitive to the strong force, is due to
the decoupling of the leptonic from the hadronic current. Limiting the
precision on extractions of the couplings $|V_\mathrm{ub}|$ and $|V_\mathrm{cb}|$ are our knowledge
of the form factors parameterizing the hadronic current. Form factors from
$B$ and $D$ meson semileptonic decay can and have been calculated using
lattice QCD techniques~\cite{ref:unquenched,ref:unquenched2,ref:quenched} whilst heavy quark symmetry relates the two form
factors~\cite{ref:theory}. Measurements of these decays are required to confront the
theoretical predictions. In this Letter, we report measurements of the
absolute rate and form factors of $D^0 \to K^- l^+ \nu_l$ and $D^0 \to \pi^- l^+
\nu_l$ ($l=e,\mu$), which have also been recently investigated by
CLEO~\cite{ref:cleonew,ref:cleoff}, BES~\cite{HQL4} and FOCUS~\cite{ref:focus}. The measurement of $D^0 \to \pi^- \mu^+ \nu_\mu$ is the first of its kind;  furthermore, measurements of the form
factor distributions $f_+(q^2)$, where $q^2$ is the invariant mass of the
lepton pair, are substantially improved by using a novel reconstruction
method with better $q^2$ resolution than in previous experiments.

Our analysis is based on
data collected by the Belle detector~\cite{Belle_det} at the asymmetric-energy KEKB storage rings~\cite{kekb} with a 
center of mass (CM) energy of 10.58 GeV ($\Upsilon(4S)$) and 60 MeV below, corresponding to a total integrated luminosity of 282 
fb$^{-1}$.
The Belle detector is a large-solid-angle magnetic spectrometer that consists of a silicon vertex detector, a 50-layer central drift chamber (CDC), an array of aerogel threshold Cherenkov counters (ACC), a barrel-like arrangement of time-of-flight scintillation counters (TOF), and an electromagnetic calorimeter comprised of CsI(Tl) crystals (ECL) located inside a superconducting solenoid coil that provides a 1.5 T magnetic field. An iron flux return located outside of the coil is instrumented to detect $K^0_L$ mesons and to identify muons.


To achieve good resolution in the neutrino momentum and $q^2$, we tag the $D^0$ by fully reconstructing the remainder of the event.
We seek events of the type $e^+e^-\rightarrow D_\mathrm{tag}^{(*)}D_\mathrm{sig}^{*-}X$ $\{D_\mathrm{sig}^{*-}\rightarrow \bar D_\mathrm{sig}^0\pi^-\}$, where $X$ may include additional $\pi^\pm$, $\pi^0$, or $K^\pm$ mesons (inclusion of charge-conjugate states is implied throughout this report).
Each candidate is assembled from a fully reconstructed ``tag-side'' charm meson ($D_\mathrm{tag}^{(*)}$) and additional particles ($X$), with the requirement that the combination be kinematically consistent with $e^+e^-\rightarrow D_\mathrm{tag}^{(*)}D_\mathrm{sig}^{*-}X$.
To the $D_\mathrm{tag}^{(*)}X$ is added a charged pion that is kinematically consistent with $\pi^-_s$ from $D_\mathrm{sig}^{*-}\rightarrow\bar D^0_\mathrm{sig}\pi_s^-$.
Candidate $D_\mathrm{tag}^{(*)}X\pi_s^-$ combinations passing the analysis criteria thus provide a tag of $\bar D_\mathrm{sig}^{0}$ and its momentum without having detected any of its decay products.
The decay $\bar D_\mathrm{sig}^{0}\rightarrow K^+(\pi^+)\ell^-\bar\nu$ may thus be reconstructed with the neutrino momentum fully constrained.

\begin{figure}
  \begin{center}
    \includegraphics[width=8.0cm,bb=0 300 500 520]{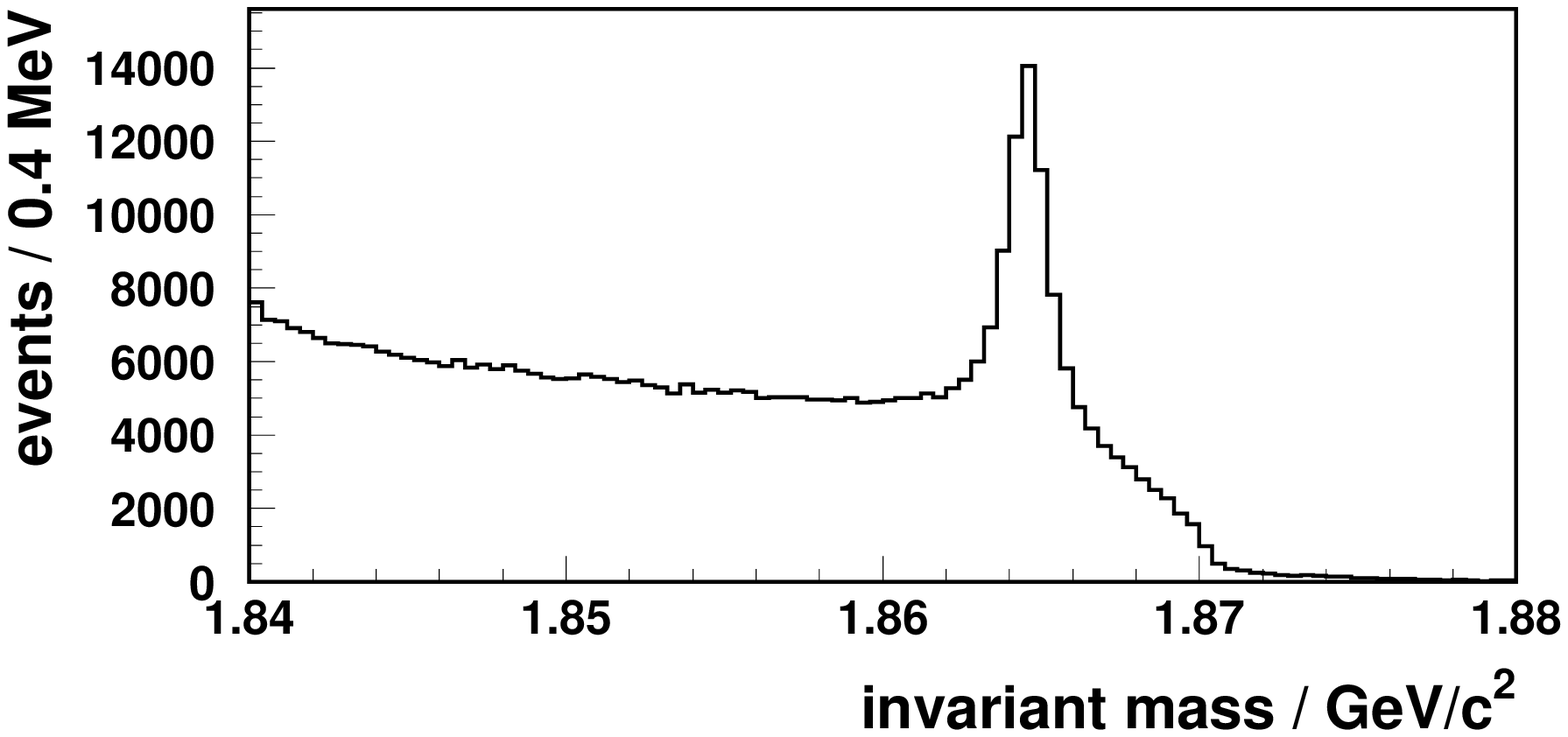}
  \end{center}
  \caption{Mass spectrum of $\bar D_\mathrm{sig}^0$ candidates.} \label{fig:n1}
\end{figure}

The $D_\mathrm{tag}^{(*)}$ is reconstructed in the modes $D^{*+}\rightarrow D^0\pi^+, D^+\pi^0$ and $D^{*0}\rightarrow D^0\pi^0, D^0\gamma$, with $D^{+/0}\rightarrow K^-(\mathrm{n}\pi)^{++/+}$ \{$\mathrm{n}=1,2,3$\}.  
Each $D_\mathrm{tag}$ and $D_\mathrm{tag}^*$ candidate is subjected to a mass-constrained vertex fit to improve the momentum resolution.
We require a successful fit of each $D_\mathrm{tag}$ candidate; furthermore, if this candidate is a daughter of a successfully fitted $D_\mathrm{tag}^*$ candidate, the event is treated as $D_\mathrm{tag}^{*}D_\mathrm{sig}^{*-}X$, otherwise it proceeds as $D_\mathrm{tag}D_\mathrm{sig}^{*-}X$.
The candidate $X$ is formed from combinations of unassigned $\pi$ and $K^+K^-$ pairs, conserving total electric charge.
The 4-momentum of $D_\mathrm{sig}^{*-}$ is found by energy-momentum conservation, assuming a $D_\mathrm{tag}^{(*)}D_\mathrm{sig}^{*-}X$ event.
Its resolution is improved by subjecting it to a fit of the $X$ tracks and the $D_\mathrm{tag}^{(*)}$ momentum, constrained to originate at the run-by-run average collision point, while the invariant mass is constrained to the nominal mass of a $D^{*-}$. The candidate is rejected if the confidence level of this fit is less than 0.1\% (corresponding to $\pm3.3\sigma$ of mass resolution).
Candidates for $\pi_s^-$ are selected from among the remaining tracks, and for each the candidate $\bar D_\mathrm{sig}^0$ 4-momentum is calculated from that of the $D_\mathrm{sig}^{*-}$ and $\pi_s^-$. The corresponding invariant mass is shown in Fig. \ref{fig:n1}. The momentum is then adjusted by a kinematic fit constraining the candidate mass to that of the $D^0$. For this fit, the decay vertex of the $\bar D_\mathrm{sig}^0$ has been estimated by extrapolating from the collision point in the direction of the $\bar D_\mathrm{sig}^0$ momentum assuming the average decay length; a comparison with the result without this vertex correction showed that the corresponding systematic error is negligible.  
Again, the candidate is rejected if the confidence level of this fit is less than 0.1\%.

Background lying under the $\bar D_\mathrm{sig}^0$ mass peak (i.e. fake-$\bar D_\mathrm{sig}^0$)
is estimated using a wrong sign (WS) sample
where the tag- and signal-side $D$ candidates have the same flavor
 ($\bar D_\mathrm{tag}$ instead of $D_\mathrm{tag}$).
A MC study (including $\Upsilon(4S) \rightarrow B\overline{B}$ and
continuum ($q\bar{q}$, where $q = c$, $s$, $u$, $d$) events~\cite{ref:bellegen,ref:bellemc}) has found that this sample can
properly model the shape of background except for a small contribution from real 
$\bar D_\mathrm{sig}^0$ decays ($\approx 2\%$) from interchange between particles used for the tag due to particle misidentification. 
Background from fake $\bar D_\mathrm{sig}^0$ is subtracted normalizing this shape in a sideband region $1.84-1.85$ GeV/$c^2$. We find $56461\pm309_\mathrm{stat}\pm830_\mathrm{syst}$ signal $\bar D_\mathrm{sig}^0$ tags. The systematic
uncertainty derives from: statistics of the WS sample ($0.5\%$); subtraction
of real $\bar D_\mathrm{sig}^0$ contamination in the WS sample ($0.6\%$) and
charge correlation in the background ($2\%$). The latter was estimated with MC by comparing true right sign (RS) background with that of the WS. 

Within this sample of $\bar D_\mathrm{sig}^0$ tags, the semileptonic decay $\bar D_\mathrm{sig}^0\rightarrow K^+(\pi^+)\ell^-\bar\nu$ is reconstructed with $K^+(\pi^+)$ and $\ell^-$ candidates from among the remaining tracks.
An event is rejected if there are any remaining unassigned charged particles ($2.2\%$ [$8.3\%$] of events in kaon [pion] mode) or if the remaining unassigned neutral energy exceeds 700~MeV ($9.0\%$ [$12.0\%$] of events in kaon [pion] mode). This requirement has been optimized based on 
a comparison of the simulated and observed energy distributions; it removes a large fraction of $\bar D_\mathrm{sig}^0$ hadronic decays with one of the final state charged tracks unreconstructed. 
The neutrino candidate 4-momentum is reconstructed by energy-momentum conservation, and its invariant mass squared, $m_{\nu}^2$, is required to satisfy $|m_{\nu}^2|<0.05~{\rm GeV}^2/c^4$.

Multiple candidates still remain in one third of $\bar D_\mathrm{sig}^0$ tags, and in
about one quarter of the semileptonic sample. In these cases all
candidates are saved and given weights adding up to $1$. 
In the MC, the difference between the result of this method
and the result when only the true signal is retained, is found
to be negligible.


\begin{table*}
\caption{Yields in data, estimated backgrounds, extracted signal
  yields and branching fractions, where for the latter two, the first uncertainty is statistical
  and second is systematic; small differences in the numbers are due to rounding.}
\label{tab:events}
\footnotesize
\begin{tabular}{l||r|r|r|r|r}
\hline\hline
channel &
\multicolumn{1}{c|}{full $\overline{D}^0_\mathrm{sig}$} &
\multicolumn{1}{c|}{$K^+e^-\nu_e$} &
\multicolumn{1}{c|}{$K^+\mu^-\nu_\mu$} &
\multicolumn{1}{c|}{$\pi^+e^-\nu_e$} &
\multicolumn{1}{c}{$\pi^+\mu^-\nu_\mu$} \\
\hline
Yield &
$95250\phantom{\,\,\,\pm 000 \pm 000}$ & 
$1349\phantom{.0 \pm 00\pm 0.0}$ &  
$1333\phantom{.0 \pm 00\pm 00.0}$ &  
$152\phantom{.0 \pm 00\pm 0.0}$ &  
$141\phantom{.0 \pm 00\pm 0.0}$ \\
fake $\overline{D}^0_\mathrm{sig}$ &
$38789\phantom{\,\,\,\pm 000} \pm 830$ &
$\phantom{0}12.6\phantom{\,\,\pm 00}\pm 2.2$ &  
$\phantom{12}12.2\phantom{\pm .....}\pm \phantom{1}4.8$ &  
$\phantom{1}12.3\phantom{\pm .....}\pm 2.2$ &  
$\phantom{1}12.5\phantom{\pm .....}\pm 4.5$ \\  
semileptonic &
n/a$\phantom{0 \pm 000 \pm 000}$ & 
$\phantom{.0}6.7\phantom{\,\,\pm 00}\pm 2.6$ & 
$\phantom{12}10.0\phantom{\pm .....}\pm \phantom{1}2.5$ &  
$\phantom{1}11.7\phantom{\pm .....}\pm 1.2$ &  
$\phantom{1}12.6\phantom{\pm .....}\pm 1.9$ \\  
hadronic &
n/a$\phantom{0 \pm 000 \pm 000}$ &
$\phantom{0}11.9\phantom{\,\,\pm 00}\pm 5.6$ &  
$\phantom{12}62.1\phantom{\pm .....}\pm 23.9$ &  
$\phantom{12}1.8\phantom{\pm .....}\pm 0.7$ &  
$\phantom{12}9.7\phantom{\pm .....}\pm 3.7$ \\  
 \hline
signal &
$56461\pm 309\pm 830$ &  
$1318\phantom{.0}\pm 37\pm 7\phantom{.0}$ &  
$1249\phantom{.1}\pm 37\pm  25\phantom{.0}$ &  
$126\phantom{.0}\pm 12\pm 3\phantom{.0}$ &  
$106\phantom{.0}\pm 12\pm 6\phantom{.0}$ \\
\hline
\multicolumn{2}{l|}{} &
\multicolumn{1}{c|}{} &
\multicolumn{1}{c|}{} &
\multicolumn{1}{c|}{} &
\multicolumn{1}{c}{} \\
\multicolumn{2}{l|}{Branching Fraction ($10^{-4}$)} &
\multicolumn{1}{c|}{$345 \pm 10 \pm 19$} &
\multicolumn{1}{c|}{$345 \pm 10 \pm 21$} &
\multicolumn{1}{c|}{$27.9 \pm 2.7 \pm 1.6$} &
\multicolumn{1}{c}{$23.1 \pm 2.6 \pm 1.9$} \\
\multicolumn{2}{l|}{($e$ and $\mu$ channels, average)} &
\multicolumn{2}{c|}{$345 \pm 7 \pm 20$} &
\multicolumn{2}{c}{$25.5 \pm 1.9 \pm 1.6$} \\
\hline
\end{tabular}
\end{table*}

The contribution from fake $\bar D_\mathrm{sig}^0$ in the sample of semileptonic decay candidates is estimated using the $\bar D_\mathrm{sig}^0$ invariant mass WS shape of the $\bar D_\mathrm{sig}^0$ tag sample, normalized in the previously defined sideband region. The effect of the additional selection criteria on the background shape has been conservatively estimated, by varying these criteria, to be $15\%$ ($35\%$) for $e$ ($\mu$) modes.

There are also backgrounds from semileptonic decays with either an
incorrectly identified meson or where additional mesons are lost in
reconstruction. These backgrounds are highly suppressed by the good
neutrino mass resolution. For $\bar D_\mathrm{sig}^0 \rightarrow \pi^+ \ell^- \nu$ the most significant
background is $\bar D^0 \rightarrow K^+ \ell^- \nu$
amounting to $6\%-8\%$ of the total yield. It
was estimated using the reconstructed $\bar D^0 \to K^+\ell^-\nu$ decays in data,
reweighted with the (independently measured) probability of kaons to fake pions. Smaller backgrounds from $\bar D^0 \rightarrow K^{*+} \ell^- \nu$ and
$\bar D^0 \rightarrow \rho^+ \ell^- \nu$ decays amounting to $0.8\%-0.9\%$ were measured by normalizing MC to
data in the upper sideband region $m^2_\nu>0.3\ \mbox{GeV}^2/c^4$, which is dominated by these channels.  For $\bar D_\mathrm{sig}^0 \rightarrow K^+ \ell^- \nu$, decays
of $\bar D^0 \rightarrow K^{*+} \ell^- \nu$ contribute at the level of $0.5\%-0.8\%$, measured using a sideband evaluation as described above, while background from
$\bar D^0 \to \pi^+ \ell^- \nu$ and $\bar D^0 \to \rho^+ \ell^- \nu$ was found to be negligible ($< 0.07\%$ of the total yield).
Systematic uncertainties are assigned due to the following: MC statistics, which
dominates overall (and according to channel ranges between $10\%-40\%$ of the
background); fake rate uncertainties ($3\%-4\%$); and uncertainty on branching
fractions of $\bar D_\mathrm{sig}^0 \to K^+/\rho^+ \ell^- \nu$ ($\approx 1\%$).

The last source of background originates from $\bar D_\mathrm{sig}^0$ decays to hadrons, where a
hadron is mis-identified as a lepton.
It is measured with an opposite sign (OS) sample, where the lepton charge is opposite to that of the $D^{*-}_\mathrm{sig}$ slow pion. 
Note that the
signal is extracted from the same sign (SS) sample.
In contrast to the SS sample, the OS sample has no
signal or semileptonic backgrounds; fake $\bar D^0_\mathrm{sig}$ are subtracted in the same manner described previously.
Assigning well identified pion and kaon tracks a lepton mass, we construct pure background $m_\nu^2$ distributions in both SS and OS, which we label $f_m^\mathrm{SS}$ and $f_m^\mathrm{OS}$, $m=K,\pi$.
A fit of the weights $a_K$ and $a_\pi$ of the components $f_K^\mathrm{OS}$
and $f_\pi^\mathrm{OS}$ in the $m_\nu^2$ distribution of the OS data
sample is performed, and the hadronic background in the SS data
sample is calculated as $(a_K f_K^\mathrm{SS} + a_\pi f_\pi^\mathrm{SS})$, utilizing the fact that the hadron misidentification rate does not depend on the charge correlation defining SS and OS. The method has been
validated using MC samples. As the muon fake rate is about an order of
magnitude larger than that for electrons, this background is much more
significant for muon modes. Systematic uncertainties are assigned based on
the bias of the method as studied in MC ($11\%-35\%$), and parameter errors from the fit ($16\%-35\%$).

The signal yields and estimated backgrounds are summarized in the upper part of Table~\ref{tab:events}. 
Efficiencies depend strongly on $n_X$, defined as the number of $\pi^{\pm(0)}$ and $K^\pm$ mesons assigned to $X$ (in $e^+e^-\rightarrow D_\mathrm{tag}^{(*)}D_\mathrm{sig}^{*-}X$), and are determined with MC; typical ratios are
$\epsilon_{h\ell\nu}/\epsilon_{\bar D_\mathrm{sig}^0} \approx 70\%$.
As the observed $n_X$ distribution in the data is slightly
different from that simulated, we reweighted the
simulated efficiencies, amounting to a $(+1.9\pm3.9)\%$ change in the efficiency
 correction; the corresponding uncertainty is accounted for in the systematic
 error.  No other biases due to $n_X$-dependent effects were found.

Normalizing to the total number of $\bar D_\mathrm{sig}^0$ tags, the absolute branching fractions summarized in the lower part of Table \ref{tab:events} are obtained; systematics are dominated by the absolute normalization.
The results for electron decay modes are in good agreement with those from~\cite{ref:cleonew,HQL4}, and the measured $\bar D_\mathrm{sig}^0 \to \pi^+ e^- \nu$ branching fraction confirms the prediction of~\cite{ref:fajfer}.
The results for the muon modes are in agreement with the ratio given in~\cite{ref:focus2}. 

More information about the semileptonic decays is obtained by studying the differential decay width $d\Gamma/dq^2$.
The resolution in $q^2$ of semileptonic decays is found to be $\sigma_{q^2}=0.0145\pm 0.0007_\mathrm{stat}$~GeV$^2/c^2$ in MC signal events.
This is much smaller than statistically reasonable bin widths, which have been chosen as $0.067\ (0.3)\ \mbox{GeV}^2/c^2$ for kaon (pion) modes, and hence 
no unfolding is necessary.
Bias in the measurement of $q^2$ that may arise due to events where 
the lepton and meson are interchanged, a double mis-assignment, was 
checked with candidate $\bar D_\mathrm{sig}^0 \to K^+\ell^-\nu$ events and found to be negligible.  
The differential decay width is bin-by-bin background subtracted and efficiency
corrected, using the same methods described previously. 

In the theoretical description, the differential decay width is dominated by 
the form factor $f_+(q^2)$~\cite{ref:2a}.
Up to order $m_{\mathrm \ell}^2$ it is given by
\begin{eqnarray}
\frac{d\Gamma^{K(\pi)}}{dq^2} & = & \frac{G_F^2|V_{cs(d)}|^2}{24\pi^3}{|f_+^{K(\pi)}(q^2)|^2}p_{\mathrm K(\pi)}^3
\label{equ:ff}
\end{eqnarray}
where $p_{\mathrm K(\pi)}$ is the magnitude  of the meson 3-momentum in the $\bar D_\mathrm{sig}^0$ rest frame \footnote{For our results, we use the full formula of Ref.~\cite{ref:2a}, including $m^2_l$ terms.}.
These form factors have been calculated recently in unquenched lattice 
QCD~\cite{ref:unquenched,ref:unquenched2}. 
In the {\it modified pole model}~\cite{ref:2}, the form factor $f_+$ is described as
\begin{eqnarray}
  f_+(q^2) &= & \frac{f_+(0)}{(1-q^2/m_{\mathrm{pole}}^2)(1-\alpha_p q^2/m_{\mathrm{pole}}^2)},
\end{eqnarray}
with the pole masses predicted as
$m(D_s^*) = 2.11$ GeV/$c^2$ (for $\bar D_\mathrm{sig}^0 \rightarrow K^+ \ell^- \nu$) and
$m(D^*) = 2.01$ GeV/$c^2$ (for $\bar D_\mathrm{sig}^0 \rightarrow \pi^+ \ell^- \nu$). 
Setting $\alpha_p = 0$ leads to the {\it simple pole model}~\cite{ref:2a}.
The ISGW2-model~\cite{ref:3} predicts
\begin{eqnarray}
f_+(q^2) &= & \frac{f_+(0)(1+\alpha_I q^2_{max})^2}{(1-\alpha_I (q^2-q^2_{max}))^2}
 \label{equ:isgw2}
\end{eqnarray}
 where $q^2_{max}$ is the kinematical limit of $q^2$ and $\alpha_I$ is a 
parameter of the model, calculated for the $K$-mode as $\alpha_I(K) = 0.47$ GeV$^{-2}c^2$.

The measured $q^2$ distribution
is fitted with $2$ free parameters to the predicted 
differential decay width $d\Gamma/dq^2$ of the different models
with $f_+(0)$ being one of the parameters, and $m_{\mathrm{pole}}$, $\alpha_p$ or $\alpha_I$ respectively the other. Binning effects are accounted for by averaging the
model functions within individual $q^2$ bins.
The fit to the simple pole model yields
$m_\mathrm{pole}(K^- \ell^+ \nu) = 1.82\pm0.04_\mathrm{stat}\pm0.03_\mathrm{syst}$~GeV/$c^2$ ($\chi^2/\mbox{ndf}=34/28$)
and $m_\mathrm{pole}(\pi^- \ell^+ \nu) = 1.97\pm0.08_\mathrm{stat}\pm0.04_\mathrm{syst}$~GeV/$c^2$ ($\chi^2/\mbox{ndf}=6.2/10$),
in agreement with results from 
CLEO~\cite{ref:cleoff} and FOCUS~\cite{ref:focus}.
While the pole mass for the $\pi \ell \nu$ decay agrees within errors with the predicted value, $m(D^*)$, the more accurate fit of $m_{\mathrm{pole}}(K \ell \nu)$ is 
several standard deviations below $m(D_s^*)$.
In the modified pole model, $\alpha_p$ describes this deviation of the real poles from the $m(D_\mathrm{(s)}^*)$ masses. Fixing these masses to their known experimental values, a fit of  $\alpha_p$ yields 
$\alpha_p (D^0 \to K^- \ell^+ \nu) = 0.52 \pm 0.08_\mathrm{stat} \pm 0.06_\mathrm{syst}$ ($\chi^2/\mbox{ndf}=31/28$) and  
$\alpha_p (D^0 \to \pi^- \ell^+ \nu) = 0.10 \pm 0.21_\mathrm{stat}  \pm 0.10_\mathrm{syst}$ ($\chi^2/\mbox{ndf}=6.4/10$).
Finally, a fit of the parameter $\alpha_I$ in the ISGW2 model yields
$\alpha_I (D^0 \to K^- e^+ \nu) = 0.51 \pm 0.03_\mathrm{stat}  \pm 0.03_\mathrm{syst}  \mbox{ GeV}^{-2}c^2$ ($\chi^2/\mbox{ndf}=33/28$)  and 
$\alpha_I (D^0 \to \pi^- e^+ \nu) = 0.60 \pm 0.10_\mathrm{stat}  \pm 0.09_\mathrm{syst}   \mbox{ GeV}^{-2}c^2$ ($\chi^2/\mbox{ndf}=7.0/10$) .
Systematic uncertainties were studied using a toy MC where the exact
simple pole model distributions for signal were randomly smeared
according to the Gaussian errors found in the data. The fit
reproduces the input pole masses without any significant bias; a
shift of $1.2\%$ ($0.3 \sigma_{\mathrm{stat}}$) observed in the pion mode was
included in the systematic error.
The subtracted background levels, which cause a correlation between $q^2$ bins, were also varied in this toy MC.

The fitted values for $f_+^{K,\pi}(0)$ vary little for the different fits, for the modified pole model the results are
$f_+^K(0) = 0.695 \pm 0.007_{\mathrm{stat}} \pm 0.022_{\mathrm{syst}}$ and $f_+^\pi(0) = 0.624 \pm 0.020_{\mathrm{stat}} \pm 0.030_{\mathrm{syst}}$; for the ratio (refitted without correlations due to normalization) we find
\begin{eqnarray}
\frac {f_+^\pi(0)^2|V_{cd}|^2} {f_+^K(0)^2|V_{cs}|^2} & = & 0.042\pm 0.003_{\mathrm{stat}}\pm 0.003_{\mathrm{syst}} \label{equ:ratio}
\end{eqnarray}
which is consistent within errors with the model-independent result using only the data in the first $\pi\ell\nu$ $q^2$ bin ($q^2<0.3\mathrm{\ GeV}^2/c^2$). A recent theoretical prediction for the ratio~\cite{ref:unquenched} is $0.040\pm0.002_\mathrm{stat}\pm0.005_\mathrm{syst}$.
Our result~(\ref{equ:ratio}) is in good agreement with those from CLEO~\cite{ref:cleoff} and FOCUS~\cite{ref:focus2}, which measure slightly lower values.

\begin{figure}[t]
  \begin{center}
        \includegraphics[width=8.0cm]{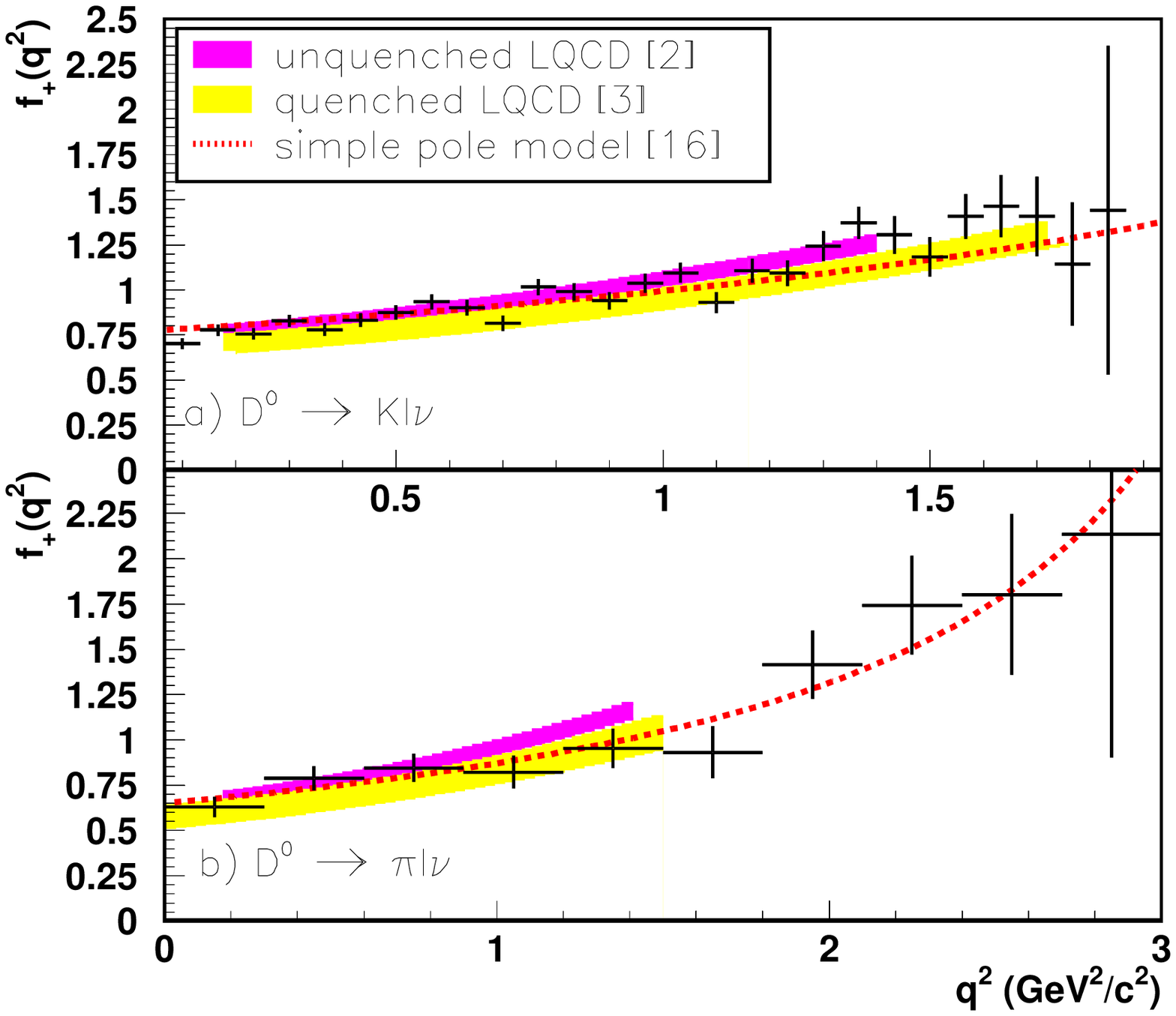} 
  \end{center}
  \caption{  
Form factors for (a) $D^0 \rightarrow K^- \ell^+ \nu$, in $q^2$ bins of $0.067\ \mbox{GeV}^2/c^2$ and
(b) $D^0 \rightarrow \pi^- \ell^+ \nu$, in $q^2$ bins of $0.3\ \mbox{GeV}^2/c^2$.  Overlaid are the
predictions of the simple pole model using the physical pole mass
(dashed), and a quenched (\cite{ref:quenched}, light gray) and unquenched (\cite{ref:unquenched2}, dark grey)
LQCD calculation.  Each LQCD curve is  obtained by fitting a parabola
to values calculated at specific $q^2$ points. The
shaded band reflects the theoretical uncertainty (without the BK-ansatz error for \cite{ref:unquenched2}) and is shown within the range
of $q^2$ for which calculations are reported.}
 \label{fig:X}
\end{figure}
The measured form factors $f_+^{K,\pi}(q^2)$ are shown in Figure \ref{fig:X} with predictions
of the simple pole model, unquenched~\cite{ref:unquenched2} and quenched~\cite{ref:quenched} LQCD.  To
obtain a continuous curve for $f_+$ from the LQCD values reported  at
discrete $q^2$ points, the values were fitted by a parabola, which is
found to fit well within the stated theoretical errors and is not
associated with any specific model.  
To quantify the degree of
agreement, we calculate $\chi^2$/ndf between our measurement and the interpolated LQCD curve within the $q^2$ range for which LQCD predictions are made.
We find a $\chi^2/\mbox{ndf}$ of $28/18$ ($34/23$) for the kaon modes and $9.8/5$ ($3.4/5$) for the pion modes; correlations induced by the fit of the calculated $q^2$ points to a parabola have been considered. 

In conclusion, our measurement of the semileptonic $D^0 \to K(\pi)\ell\nu$ decays yields absolute branching fractions in agreement with other new measurements, and the first measurement of the absolute branching fraction, $\mathcal{B}(D^0\rightarrow\pi^- \mu^+\nu)$. The good $q^2$ resolution results in substantially improved measurements of $D^0\rightarrow K^-(\pi^-)e^+\nu$ $f_+(q^2)$-distributions.

We thank the KEKB group for excellent operation of the
accelerator, the KEK cryogenics group for efficient solenoid
operations, and the KEK computer group and
the NII for valuable computing and Super-SINET network
support.  We acknowledge support from MEXT and JSPS (Japan);
ARC and DEST (Australia); NSFC and KIP of CAS (contract No.~10575109 and IHEP-U-503, China); DST (India); the BK21 program of MOEHRD, and the
CHEP SRC and BR (grant No. R01-2005-000-10089-0) programs of
KOSEF (Korea); KBN (contract No.~2P03B 01324, Poland); MIST
(Russia); ARRS (Slovenia);  SNSF (Switzerland); NSC and MOE
(Taiwan); and DOE (USA).

\end{document}